\documentclass[amsmath,amssymb,pra,twocolumn,superscriptaddress]{revtex4-2}

\usepackage{graphicx}
\usepackage{dcolumn}
\usepackage{bm}
\usepackage[colorlinks,linkcolor=blue,anchorcolor=blue,citecolor=blue,]{hyperref}
\usepackage{xcolor}

\begin{document}

\title{Rotational level spacings in HD from vibrational saturation spectroscopy}

\author{F. M. J. Cozijn}
\affiliation{Department of Physics and Astronomy, LaserLab, Vrije Universiteit\\
De Boelelaan 1081, 1081 HV Amsterdam, The Netherlands}

\author{M. L. Diouf}
\affiliation{Department of Physics and Astronomy, LaserLab, Vrije Universiteit\\
De Boelelaan 1081, 1081 HV Amsterdam, The Netherlands}

\author{V. Hermann}
\affiliation{Tritium Laboratory Karlsruhe, Institute of Astroparticle Physics,
Karlsruhe Institute of Technology\\
Hermann-von-Helmholtz-Platz 1, 76344 Eggenstein-Leopoldshafen, Germany}

\author{E. J. Salumbides}
\affiliation{Department of Physics and Astronomy, LaserLab, Vrije Universiteit\\
De Boelelaan 1081, 1081 HV Amsterdam, The Netherlands}

\author{M. Schlösser}
\affiliation{Tritium Laboratory Karlsruhe, Institute of Astroparticle Physics,
Karlsruhe Institute of Technology\\
Hermann-von-Helmholtz-Platz 1, 76344 Eggenstein-Leopoldshafen, Germany}

\author{W. Ubachs}
\affiliation{Department of Physics and Astronomy, LaserLab, Vrije Universiteit\\
De Boelelaan 1081, 1081 HV Amsterdam, The Netherlands}
 \email{w.m.g.ubachs@vu.nl}

\date{\today}

\begin{abstract}
The R(1), R(3) and P(3) ro-vibrational transitions in the (2-0) overtone band of the HD molecule are measured in Doppler-free saturation using the technique of NICE-OHMS spectroscopy.
For the P(3) line, hitherto not observed in saturation, we report a frequency of $203\,821\,936\,805\,(60)$ kHz.
The dispersive line shapes observed in the three spectra show strong correlations, allowing for extraction of accurate information on rotational level spacings.
This leads to level spacings of $\Delta_{(J=3)-(J=1)}= 13\,283\,245\,098\,(30)$ kHz in the $v=0$ ground state, and $\Delta_{(J=4)-(J=2)}= 16\,882\,368\,179\,(20)$ kHz in the $v=2$ excited vibration in HD.
These results show that experimental values for the rotational spacings are consistently larger than those obtained with advanced ab initio theoretical calculations at 1.5$\sigma$, where the uncertainty is determined by theory. The same holds for the vibrational transitions where systematic deviations of 1.7-1.9$\sigma$ are consistently found for the five lines accurately measured in the (2-0) band.

\end{abstract}

\maketitle

\section{introduction}

The spectroscopic investigation of the hydrogen molecule and its isotopologues has
played a crucial role in the advancement of quantum mechanics in the molecular domain.
The HD isotopologue, observed via its vacuum ultraviolet spectrum immediately after its production and purification~\cite{Jeppesen1934,Urey1935}, undergoes breaking of inversion symmetry, also referred to as $g$ - $u$ symmetry breaking, giving rise to spectroscopic phenomena that are not observed in the homo-nuclear species H$_2$ and D$_2$~\cite{Lange2002b}.
One of the special features is the occurrence of a dipole-allowed absorption spectrum in HD in connection the small dipole moment arising from a charge asymmetry in the molecule.
Wick was the first to calculate the intensity of this dipole-allowed vibrational spectrum~\cite{Wick1935} and Herzberg  first observed overtone lines combining high-pressure cells with multi-pass absorption~\cite{Herzberg1950}.
The vibrational spectra of the fundamental~\cite{Rich1982} and overtone bands~\cite{Durie1960} were later investigated in more detail and at higher accuracy.
In the past decade cavity-enhanced techniques in combination with frequency-comb calibration were employed to investigate the spectrum of the (2-0) band of HD~\cite{Kassi2011}.
For a literature compilation of the vibrational spectra of HD we refer to Ref.~\cite{Vasilchenko2016}.
The pure rotational spectroscopy of HD, also connected to the dipole moment, was first probed by Trefler and Gush~\cite{Trefler1968}, while later more precise spectroscopic measurements were performed~\cite{Essenwanger1984,Evenson1988,Ulivi1991,Drouin2011}.

Recently, saturation spectroscopy of R-lines in the (2-0) overtone band was demonstrated~\cite{Tao2018,Cozijn2018}, to yield linewidths much narrower than in the Doppler-broadened spectroscopies performed previously.
These studies led to accuracies at the 20 kHz level, but the modeling of observed lineshapes appeared to be a limiting factor.
In the Amsterdam laboratory subsequent measurements with the NICE-OHMS (Noise-Immune Cavity-Enhanced Optical-Heterodyne Molecular Spectroscopy) technique were carried out and the dispersive-like line shape was interpreted in terms of an Optical Bloch equation (OBE) model including underlying hyperfine structure and crossover resonances in the saturation spectrum~\cite{Diouf2019,Diouf2020}.
The Hefei group extended their studies on the observation of a dispersive line shape, which was interpreted as a Fano line shape~\cite{Hua2020}.
The uncertainties associated with the observed lineshape remain to be a dominating factor and hinder full exploitation of the extreme resolution of the saturation technique, and the determination of transition frequencies at the highest accuracy.
The molecular-beam double-resonance study of the R(0) line of the (1-0) fundamental of HD by Fast and Meek~\cite{Fast2020} does not suffer from this short-coming, yielding a vibrational splitting in HD at the level of 13 kHz, the most accurate to date.
Vibrational splittings in the (1-0) fundamental  have also been determined via laser-precision studies in molecular beams and the measurement of combination differences in Doppler-free electronic spectra~\cite{Niu2014}.
The Caserta group applied cavity-enhanced methods for linear absorption spectroscopy of the R(1) line in (2-0). Even though the line is of GHz width an accuracy of 76 kHz is obtained through advanced modeling of the Doppler-broadened line shape~\cite{Castrillo2021,Castrillo2021b}.

\begin{figure*}
\begin{center}
\includegraphics[width=0.85\linewidth]{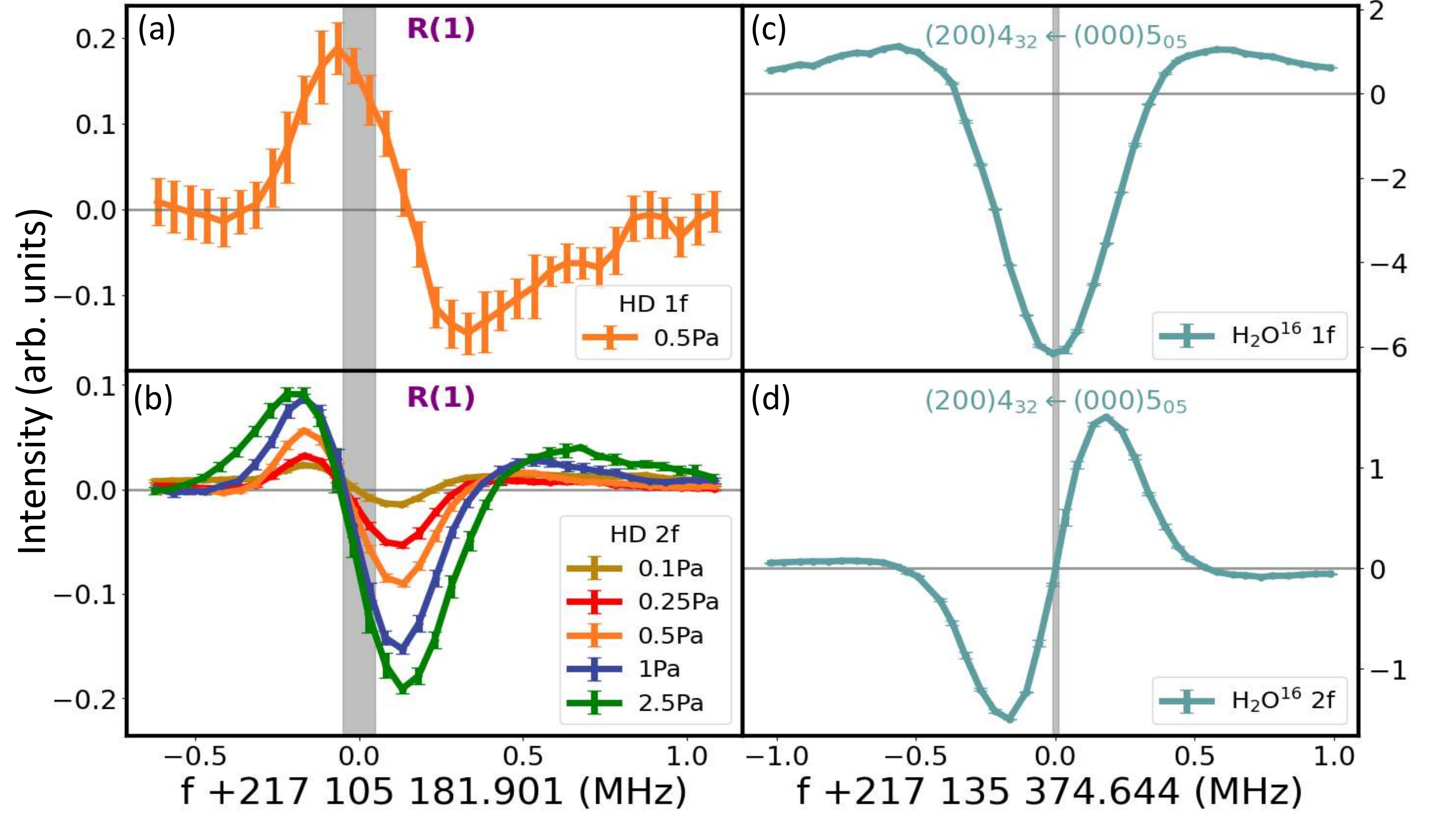}
\caption{\label{Exp-Line-Shapes}
Spectral recordings of the R(1) line in the (2-0) overtone band of HD in saturated absorption employing the NICE-OHMS technique. Pressures as indicated in units of Pa. Recordings are performed using (a) $1f$ and (b) $2f$ demodulation. A comparison is made with the recording of a water line, also at (c) $1f$ and (d) $2f$ demodulation settings. Note the strong difference in line shape, where the water line represents the generic NICE-OHMS signal shape.
The (grey) vertical bar in (a) and (b) represents the transition frequency of the R(1) line of HD as determined in a previous experimental saturation study via modelling of the spectral line shape based on underlying hyperfine structure and including cross-over resonances~\cite{Diouf2019}.
The 0.0 value represents a frequency of 217\,105\,181.901 (0.050) MHz for HD and 217\,135\,374.644 (0.005) MHz for the H$_2$O$^{16}$ line.
}
\end{center}
\end{figure*}

The goals of precision spectroscopy on the hydrogen isotopologues have surpassed the targets of molecular physics.
These smallest neutral molecular species have become benchmark systems for probing physics beyond the Standard Model~\cite{Ubachs2016}, searching for fifth forces of various nature~\cite{Salumbides2013,Ledbetter2013} and for higher dimensions~\cite{Salumbides2015b}.
The searches for new physics depend on the availability of highly accurate ab initio computations of the level structure of the hydrogen molecules.
In the past decade the boundaries in this area have been pushed, and currently highly accurate level energies  are produced in 4-particle fully variational calculations of the relativistic motion in the molecules, augmented with calculation of quantum-electrodynamic corrections up to level $m\alpha^6$~\cite{Czachorowski2018,Puchalski2018,Puchalski2019}.
The results of the ab initio calculations are available through the H2SPECTRE on-line program~\cite{SPECTRE2020}.

A comparison between the recent accurately measured vibrational transition frequencies with those computed from the H2SPECTRE code reveals a systematic offset of 1 MHz for the (1-0) band and 2 MHz for the (2-0) band.
The deviations extend very much beyond the uncertainties established in the experiments,  but remain at the level of $1.7\sigma$, when the uncertainty of the calculations is taken into account.
In order to investigate the origin of these discrepancies we target measurements, by means of saturation spectroscopy, of combination differences between vibrational lines in the (2-0) band.

\section{experiment and results}

In the experiment, the NICE-OHMS setup at the Amsterdam laboratory is used to perform saturation spectroscopy of the R(1), R(3) and P(3) lines of HD at wavelengths near $1.4\,\mu$m.
Details of the setup were described in previous papers on the spectroscopy of HD~\cite{Cozijn2018,Diouf2019,Diouf2020} and specific settings of the present experiment are similar.
A noteworthy detail is that a new diode-laser and high-reflectivity mirrors are used to reach the P(3) wavelength, while the measurements of the R(1) and R(3) lines are performed with existing components.
The essentials remain identical with an intracavity circulating power at the central carrier frequency $f_c$ of about 150 W, while the sidebands at $f_c \pm f_m$ are modulated at $f_m=404$ MHz delivering circulating powers of 2 W.
The diode-laser is locked to the high-finesse cavity (150,000) via Pound-Drever-Hall stabilization, where the cavity is locked to a Cs-clock stabilized frequency-comb laser.
This locking sequence leads to a line-narrowing of the diode laser to around 20 kHz at second time scale caused by short term vibration-noise and thermal drift of the cavity. Ultimately, the absolute frequency of the complete measurement data averages down to below kHz precision due to long term measurements of over a few hours. This stability allows for effectively averaging over multiple scans to obtain reasonable signal-to-noise levels.
Averages of around 60 scans with a total measurement time of over 10 hours were taken to record a spectrum of the weakest P(3) line.

The generic signals produced in direct NICE-OHMS spectroscopy under saturation exhibit a dispersive lineshape, as a result of the sideband frequency-modulation applied~\cite{Foltynowicz2008b,Axner2014a}.
The application of an additional low-frequency dither modulation to the cavity length (at 415 Hz) and demodulation at $1f$ by a lock-in amplifier, in principle results in a line shape taking the form of a derivative of a dispersion-like function.
Such symmetric line shapes were indeed detected for saturated lines of C$_2$H$_2$~\cite{Diouf2019} and of H$_2$O~\cite{Tobias2020} in the same setup.
In Fig.~\ref{Exp-Line-Shapes} representative spectra of an HD line, in this case the R(1) line in the (2-0) overtone band, are compared with spectral recordings of a water line. This comparison was performed for both $1f$ and $2f$ demodulation of the modulated signal.
Details of the applied modulation scheme has been presented earlier for the $1f$ signal channel \cite{Diouf2019}, but the used lock-in amplifier (Zurich Instruments HF2LI) can be expanded with an additional parallel demodulation channel allowing simultaneous measurements of the $1f$ and $2f$ signal channels.
This capability has been used in previous work to detect the $1f$ and $3f$ demodulation channels simultaneously, which resulted in resolving the hyperfine structure of H$_2^{17}$O~\cite{Melosso2021}.

\begin{figure}
\begin{center}
\includegraphics[width=1.0\linewidth]{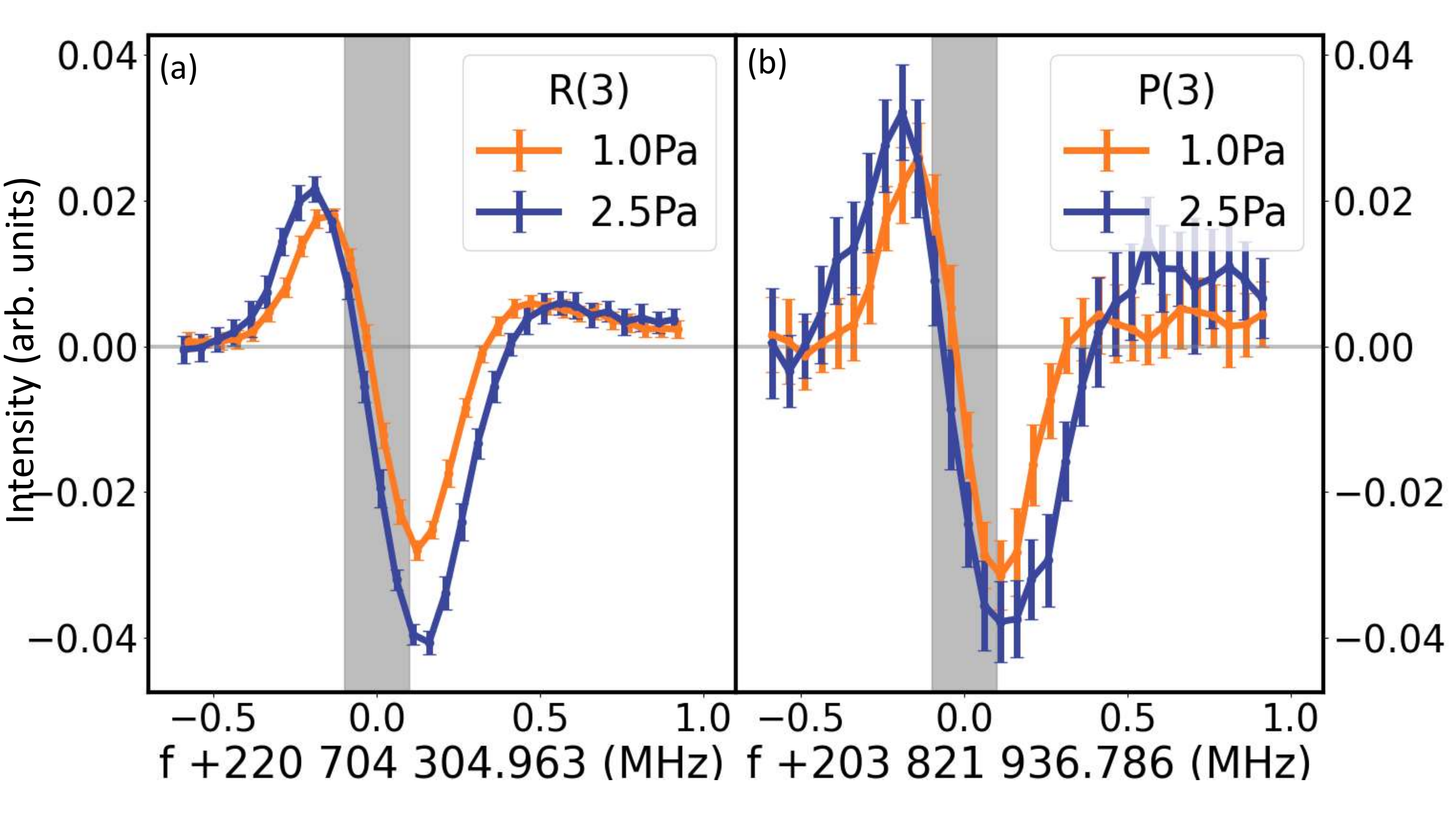}
\caption{\label{P3R3}
Recordings of saturated spectra of the R(3) and P(3) (2-0) overtone lines in HD with NICE-OHMS at $2f$-demodulation for pressures of 1 Pa and 2.5 Pa. The (grey) bars indicate the estimated spin-averaged transition frequencies and their uncertainties of 100 kHz.
}
\end{center}
\end{figure}

These spectra and the comparison between HD and H$_2$O resonances as measured in saturation demonstrate two important aspects.
Firstly, the line shape of the HD resonance in the $1f$-recording is asymmetric,  unlike the shape of the water resonance that follows the expected pattern for NICE-OHMS signals~\cite{Tobias2020}.
Similar asymmetric line shapes are observed for the R(3) and P(3) lines, as presented in Fig.~\ref{P3R3}.
This phenomenon of observing unexpected atypical line shapes  was discussed in previous papers on the saturation spectroscopy of HD, either using $1f$-demodulation in NICE-OHMS~\cite{Cozijn2018,Diouf2019}, cavity-ring-down spectroscopy~\cite{Tao2018}, or a variety of cavity-enhanced techniques~\cite{Hua2020}.
Secondly, the experimental data show that the $2f$-demodulation spectra exhibit a better signal-to-noise ratio (SNR) than the $1f$-demodulated spectra, while the $1f$ spectra display a much better SNR than the direct NICE-OHMS spectra.
For these reasons the comparisons and the detailed analyses of rotational line shifts is based on $2f$ demodulated spectra in the following.

In Fig.~\ref{Exp-Line-Shapes} the transition frequency of the R(1) line and its uncertainty are indicated by the (grey) vertical bar.
A result for the R(1) transition frequency was initially reported from a NICE-OHMS study only considering the Lamb-dip feature and fitting its line centre; this procedure resulted in a value then considered to be accurate to 20 kHz~\cite{Cozijn2018}.
However, in a competing study using cavity-ring-down spectroscopy  a strongly deviating transition frequency was reported~\cite{Tao2018}.
For this reason a systematic study was performed in which the observed complex line shape, consisting of Lamb-peak and Lamb-dip contributions, was computed via an Optical Bloch Equation (OBE) model from whicresulting ih a spin-averaged transition frequency of $217\,105\,181\,901$ kHz with an uncertainty of 50 kHz~\cite{Diouf2019}.
This is the result displayed by the  (grey) vertical bar in Fig.~\ref{Exp-Line-Shapes}.
It shows that the modelled center frequency coincides rather accurately with the Lamb-peak feature in the $1f$ NICE-OHMS signal channel, and with the zero-crossing in the $2f$ signal channel.

For  the R(3) line an accurate result was reported in a previous study~\cite{Cozijn2018}, but this was not substantiated with an explicit OBE-model computation.
The P(3) line in the (2-0) overtone band has not been reported before from a saturation experiment. Based on the finding that the line shapes of the R(1), R(3) and P(3) lines exhibit similar line shapes (this paper) it is assumed that the central spin-averaged transition frequencies are all in close proximity of the zero-crossing point in the $2f$ spectral features at the lowest pressures.
In view of this assumption the transition frequencies of R(3) and P(3) lines in the observed $2f$-spectra are extracted from the zero-pressure extrapolated $2f$ crossings, displayed in Fig.~\ref{P3R3}, with an uncertainty bar estimated conservatively at 100 kHz.
Based on the measurements performed at 1.0 and 2.5 Pa, pressure shifts of -23 kHz/Pa for R(3) and -26 kHz/Pa for P(3) are determined.
This leads to  values of $220\,704\,304\,963\,(100)$ kHz for R(3) and $203\,821\,936\,786\,(100)$ kHz for P(3).

\begin{figure}
\begin{center}
\includegraphics[width=1.0\linewidth]{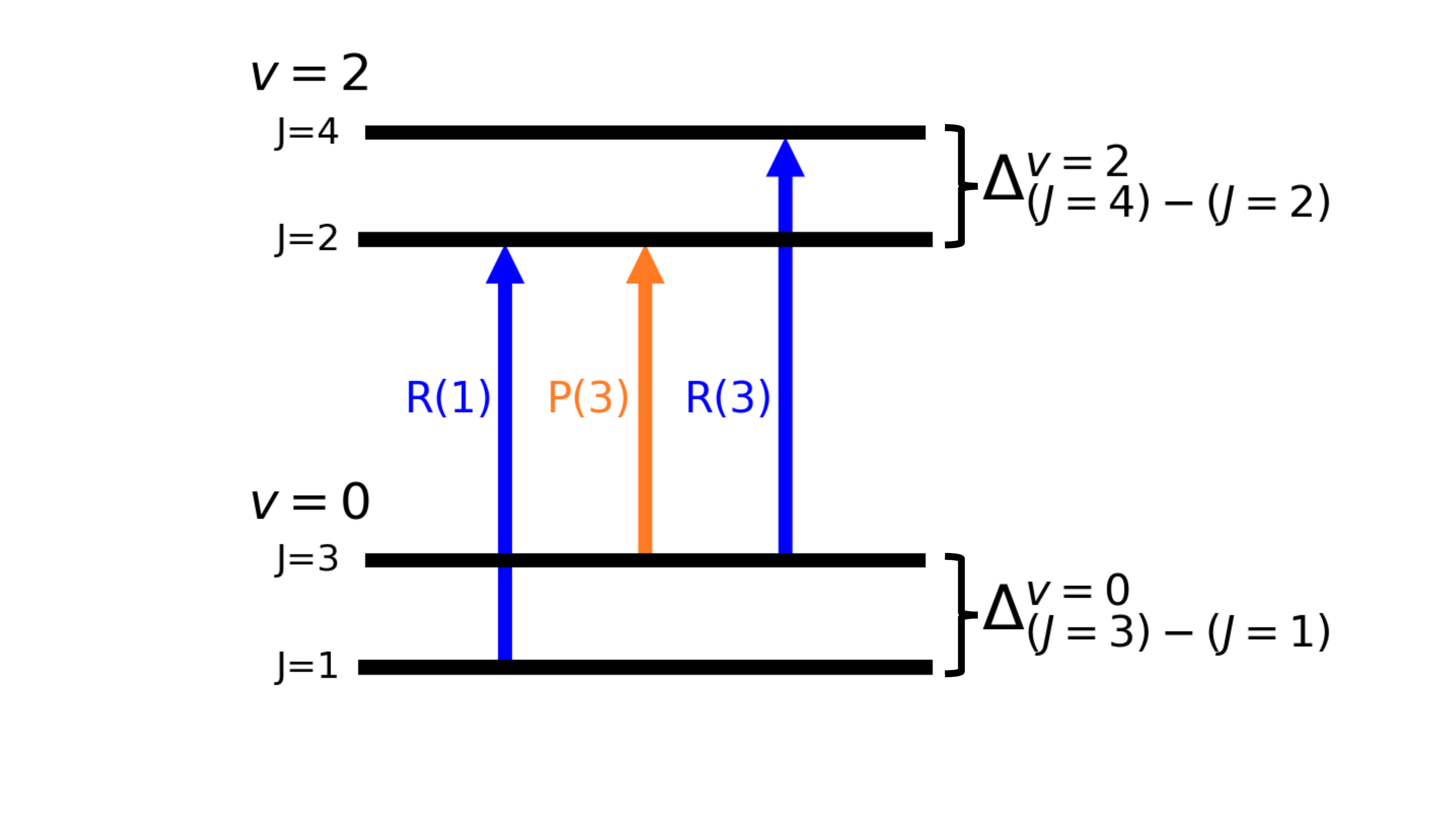}
\caption{\label{Level-scheme}
Level scheme of HD showing how the three lines R(1), R(3) and P(3) correspond to rotational level spacings in the $v=0$ ground and $v=2$ excited vibrational level.
}
\end{center}
\end{figure}

\begin{figure*}
\begin{center}
\includegraphics[width=0.85\linewidth]{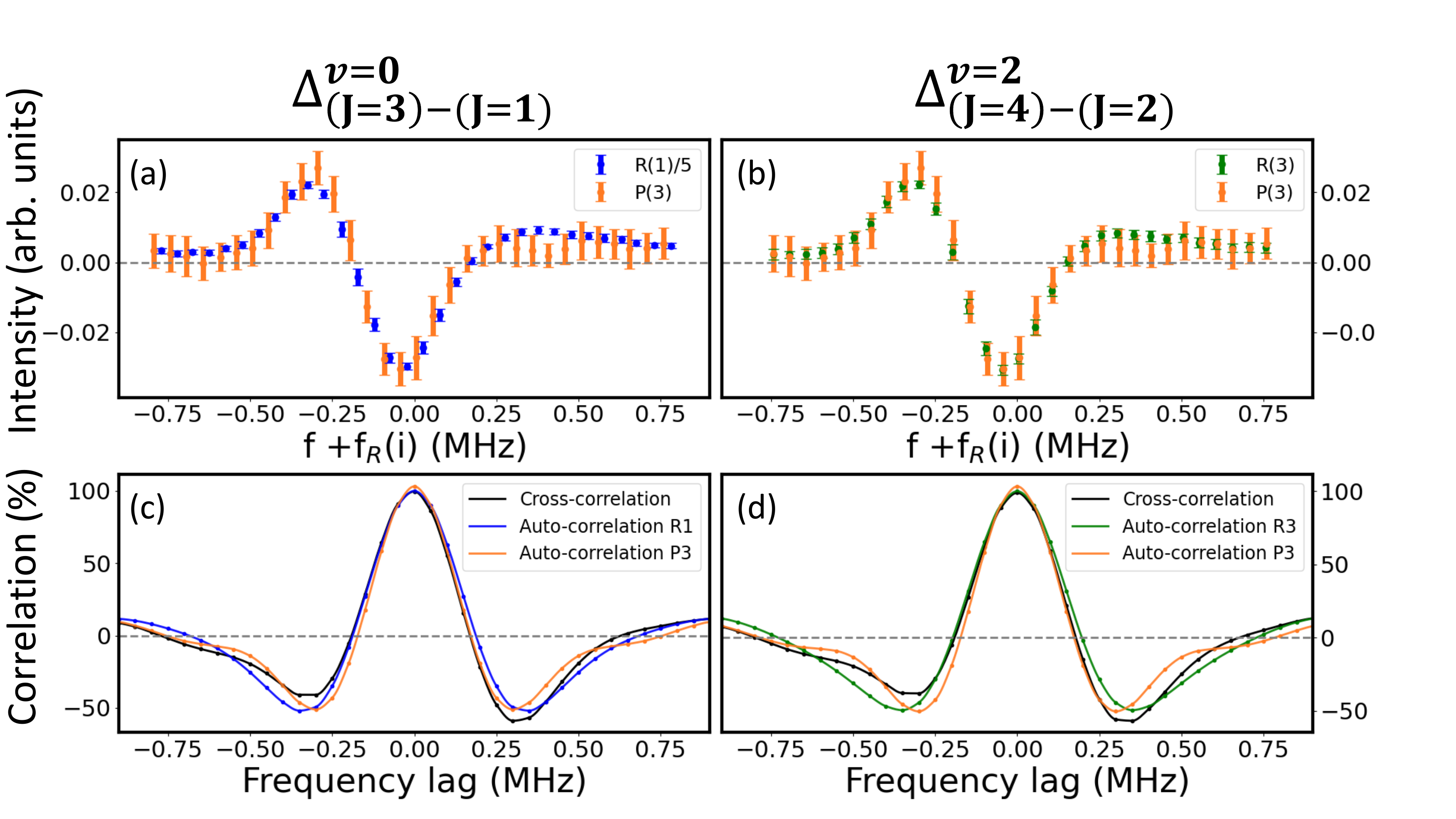}
\includegraphics[width=0.85\linewidth]{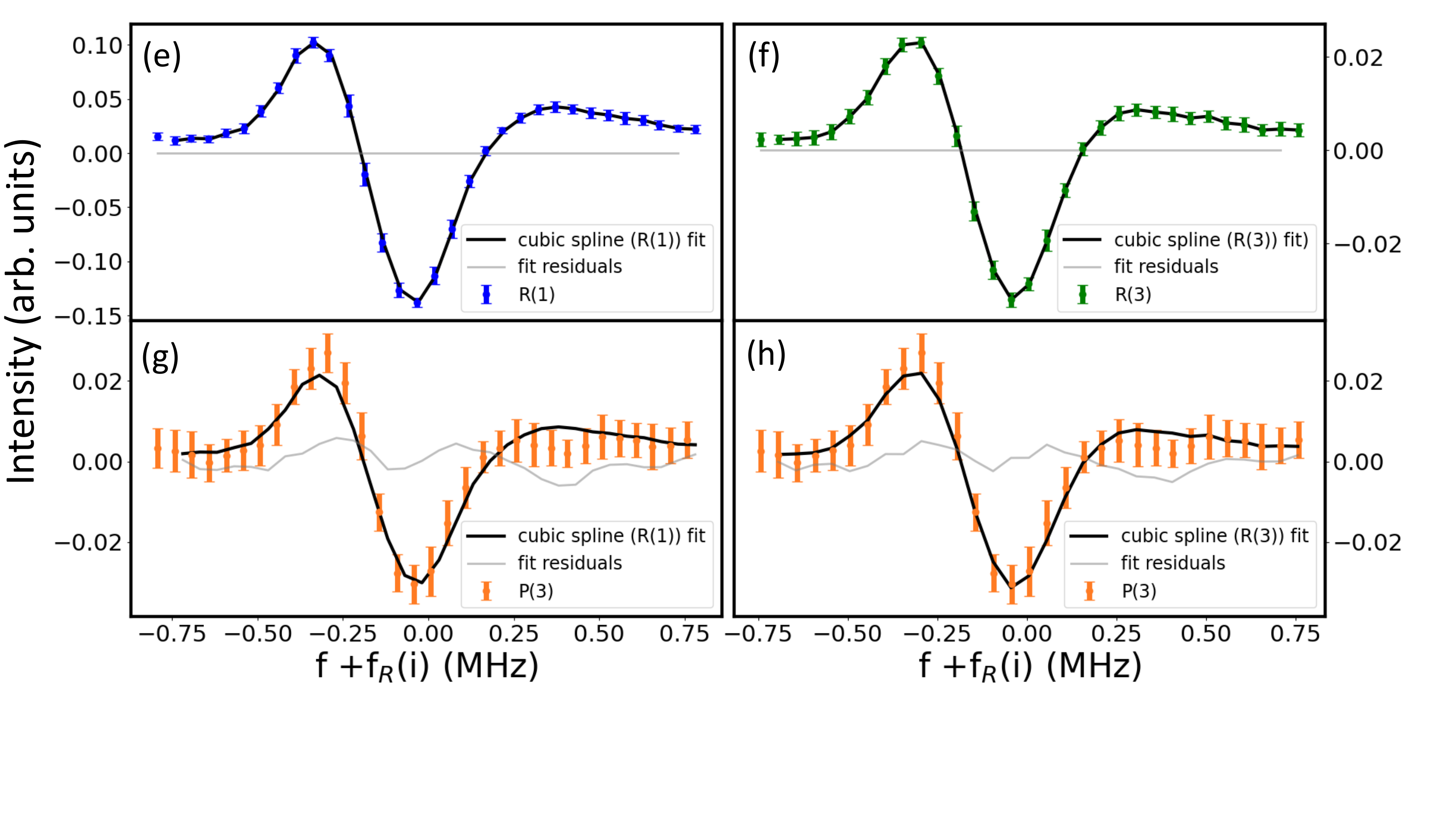}
\caption{\label{Correlations}
Comparison of the $2f$ demodulated saturation spectra for three lines in the (2-0) overtone band of HD, all recorded at a pressure of 1 Pa;
(a) Comparison for the pair R(1) and P(3); (b) Comparison for the pair R(3) and P(3).
Note that the amplitudes for the corresponding spectra are adapted to match each other.
(c) Results of auto-correlation and cross-correlation calculations for the R(1)-P(3) pair and (d) for the R(3)-P(3) pair.
In panels (e) and (f) fits are made using a cubic spline interpolation for which residuals are computed and plotted (thin grey line).
The resulting cubic spline functional form is then fitted to the line shapes of the P(3) lines in (g) and (h). Residuals of the latter are again plotted in grey.
}
\end{center}
\end{figure*}

In Figs.~\ref{Exp-Line-Shapes} and \ref{P3R3} it is documented that recorded saturation spectra for the HD resonances deviate from expected and generic line shapes in NICE-OHMS.
However, visual inspection indicates that the line shapes of the R(1), R(3) and P(3) lines are very similar.
This forms the basis for extracting information on rotational level spacings as illustrated in Fig.~\ref{Level-scheme}.

In Fig.~\ref{Correlations} recordings of $2f$-demodulated spectra for the three HD lines are compared.
For quantifying the similarities auto-correlation functions are computed as well as cross-correlation functions~\cite{bookRabiner1975}
for the pairs R(1)-P(3) and R(3)-P(3).
The resulting cross-correlation curves are found to mimic the auto-correlations in a near-perfect fashion, with a peak overlap of 95\%.
The cross-correlation computations yield values for the frequency differences $\Delta$ between the line pairs R(1)-P(3) and R(3)-P(3) representing rotational line spacings in the $v=0$ and $v=2$ levels.

Because the computation of the cross-correlation does not straightforwardly deliver an uncertainty to the values for the line spacing, the $2f$ modulated spectra for the R(1) and R(3) lines were subjected to a standard cubic spline interpolation~\cite{Hall1976-CS} to produce a functional form closely representing the data.
The cubic spline curves and plotted residuals in Fig.~\ref{Correlations}(e) and (f) show that such cubic splines indeed accurately represent the experimental data for the R(1) and R(3) lines.
These resulting functional forms, $g(f)$, were then used to fit the third P(3) line in both cases. Aside from adjustment parameters for the amplitude ($A$) and zero-level ($B$) a frequency shift term $\Delta$  between the R(1)/P(3) and R(3)/P(3) pairs was included, which results in a general fitting function $A \cdot g(f+\Delta)+B$.
The latter fits then deliver values for $\Delta$ as well as an uncertainty:
\begin{eqnarray}
\Delta_{(J=3)-(J=1)}^{v=0}= 13\,283\,245\,098\,(5) \, \rm{kHz} \nonumber \\
\Delta_{(J=4)-(J=2)}^{v=2}= 16\,882\,368\,179\,(5) \, \rm{kHz} \nonumber
\end{eqnarray}
in the $v=0$ ground and $v=2$ excited vibrations in HD.
While the results for $\Delta$ are similar as in the computations of the cross-correlations (deviations as small as 2 kHz and 3 kHz found, respectively),
the procedure based on cubic-spline fitting delivers a statistical uncertainty as small as 5 kHz for the spacing between the corresponding line pairs.

Systematic effects should be considered that contribute to the error budget for the frequency spacings between lines.
The results presented in Fig.~\ref{Correlations} pertain to measurements at a pressure of 1 Pa.
A similar analysis was performed for data sets obtained at 2.5 Pa leading to values for combination differences $\Delta$ within 2 kHz.
Indeed, the pressure effects on level spacings between rotational lines are expected to be small in view of common-mode cancelation of collisional shifts in the combined transitions.

In the previous study~\cite{Cozijn2018} it was established that power broadening does play a role in the saturated NICE-OHMS spectroscopy of HD, but power shifts are constrained to $< 1$ kHz.
Also the frequency calibration reaches kHz accuracy.

\begin{figure}
\begin{center}
\includegraphics[width=1.0\linewidth]{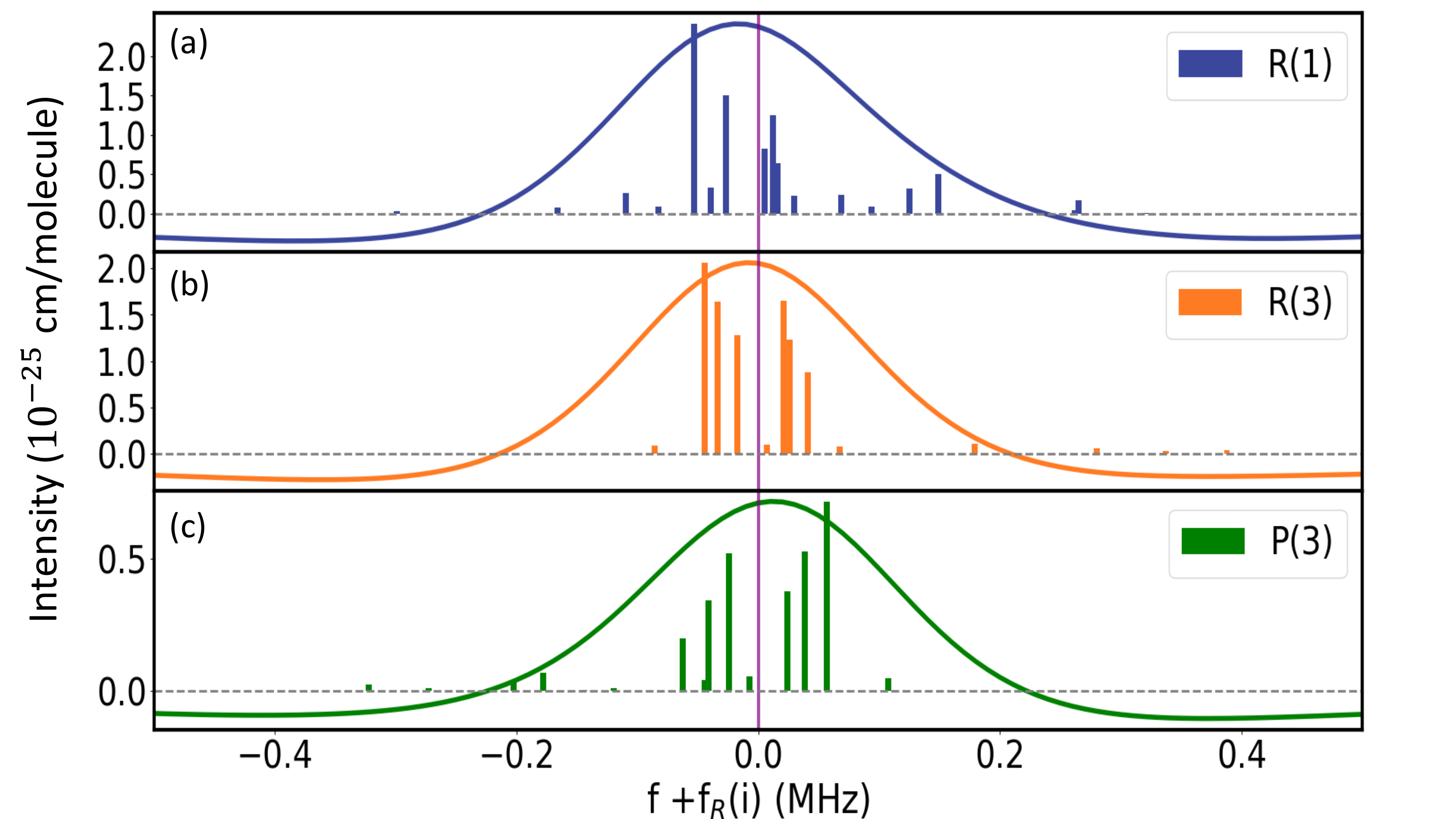}
\caption{\label{Hyperfine}
Stick spectra of hyperfine components for (a) the R(1) line, (b) the R(3) line; and (c) the P(3) line in the (2-0) overtone band of HD based on the computations of Ref.~\cite{Jozwiak2020c}.
These stick spectra were convolved with $1f$-NICE-OHMS spectral features for each component to yield the curves as plotted.
}
\end{center}
\end{figure}

The observed composite line shapes, consisting of Lamb-peaks and Lamb-dips, were in our previous studies interpreted as resulting from underlying hyperfine structure and cross-over resonances for which a quantitative model was developed based on optical Bloch equations (OBE)~\cite{Diouf2019}.
Such an analysis supported by OBE modeling was developed for the R(1) line.
In the derivation of the frequency spacings $\Delta$ it is assumed that underlying hyperfine structure does not affect the accuracy of this treatment.
To assess a possible shift caused by differences in the hyperfine structure of the three lines the underlying hyperfine structure of all three lines is compared, based on the computations of Ref.~\cite{Jozwiak2020c}.
In Fig.~\ref{Hyperfine}  stick spectra of hyperfine components in the spectra of R(1), R(3) and P(3) lines are displayed, convolved with a $1f$-demodulated NICE-OHMS function (a $1f$ derivative of a dispersive line shape) to produce a final width commensurate with the width ($\Gamma = 400$ kHz FWHM) obtained in the auto-correlation functions in Fig.~\ref{Correlations}(c,d).
For each of these convoluted functions it is computed in how far the line center deviates from the center-of-gravity of the hyperfine structure, \emph{i.e.} the deviation for  the spin-averaged frequency.
From these calculations it follows that the center frequencies are shifted from the center-of-gravity of the hyperfine structure by -20 kHz, -10 kHz and +10 kHz for the R(1), R(3) and P(3) lines respectively.
These shifts are included as systematic uncertainties in the error budget for the pure rotational line spacings.
Hence for the  P(3)/R(3) line pair a contribution of 20 kHz, and for the P(3)/R(1) pair a contribution of 30 kHz is included as a systematic uncertainty.
These estimates on the uncertainty arising from the underlying hyperfine structure make this contribution the dominant one.

Having established values for the combination differences between the pair P(3)/R(1), including uncertainty, this result can be combined with the accurate result for the transition frequency of the R(1) line based on the OBE-model~\cite{Diouf2019}, yielding the transition frequency $203\,821\,936\,805\,(60)$ kHz for the P(3) line, deviating some 19 kHz from the estimate based on the $2f$ crossing point.
Further combining the pair P(3)/R(3) then delivers a transition frequency for the R(3) line, yielding $220\,704\,304\,984\,(65)$ kHz, deviating 21 kHz from the estimates from the $2f$ crossing point.
These frequency separations, obtained via two distinct methods, are in agreement with each other within $0.3\sigma$.
The values obtained through the combination differences, considered to  be most accurate, are included in Table~\ref{tab:comparison}.

\section{Discussion and Conclusion}

\begin{table*}
\caption{\label{tab:comparison}
Comparison between experimental data on purely rotational and ro-vibrational data on transitions frequencies for HD with results  of computed results using H2SPECTRE~\cite{SPECTRE2020}.  All frequencies given in MHz. Differences are presented in terms of MHz and in terms of the combined standard deviation ($\sigma$) of experiment and theory. For the entries assigned with 'T' the deviations are entirely determined by the theoretical values from H2SPECTRE.
}
\begin{tabular}{lllcllcc}
Band	& Line & Exp. (MHz) &   Ref.  &  H2SPECTRE (MHz) &   Diff. (MHz) &   Diff. ($\sigma$) & \\
\hline
(0-0) \\
       & R(0) & 2 674 986.66 (0.15)  &  \cite{Evenson1988}   & 2 674 986.071 (0.022)  & 0.59 (0.15) &  3.9 \\
       &    & 2 674 986.094 (0.025)   &  \cite{Drouin2011}   & 2 674 986.071 (0.022)  & 0.023 (0.033) &  0.7  & (T)\\
       & R(1) &5 331 560.6 (4.8)  & \cite{Ulivi1991}  &  5 331 547.053 (0.045)  & 13.5 (4.8) & 2.8 \\
    & R(2) & 7 951 729.9 (5.1)   & \cite{Ulivi1991}  &  7 951 697.887 (0.066)  & 32 (5)   & 6.4 \\
     & R(3) & 10 518 306.8 (3.6)    & \cite{Ulivi1991}  &  10 518 308.641 (0.087)  & -1.8 (3.6)   & -0.5 \\
      & R(6) & 17 745 695.9 (9.0)  & \cite{Essenwanger1984}  & 17 745 686.540 (0.140)  & 3.8 (6.6)     & 1.0 \\
     & S(1)\footnote{Derived from a spacing between lines, or a combination difference.} & 13 283 245.098 (0.030) & Present  &  13 283 244.944 (0.110)  &    0.15 (0.11)  & 1.4 & T \\
\hline
(1-0) \\
 & Q(0)$^{\rm a}$ & 108 889 433.0 (6.6)   &  \cite{Niu2014}   & 108 889 429.2 (0.5)  & 3.8 (6.6) &  0.6 \\
 & R(0) & 111 448 815.477 (0.013)  &  \cite{Fast2020}   & 111 448 814.5 (0.6)  & 1.0 (0.6)   &  1.6   &  T    \\
 & Q(1)$^{\rm a}$ & 108 773 832.4 (6.6)   &  \cite{Niu2014}   & 108 773 828.4 (0.5)  & 4.0 (6.6) &  0.6 \\
\hline
(2-0) \\
& P(1) & 209 784 242.007 (0.020)   &  \cite{Diouf2020}   & 209 784 240.1 (1.0)  & 1.9 (1.1)  &  1.7 & T\\
 & P(3) & 203 821 936.805 (0.060)  &  Present\footnote{Results from vibrational transitions in the present study.}   & 203 821 935.0 (1.0)  & 1.8 (1.0) & 1.8  & T\\
 & R(1) & 217 105 181.901 (0.050)  &  \cite{Diouf2019}   & 217 105 180.0 (1.1)  & 1.9 (1.1)  &  1.7 & T\\
 &  & 217 105 182.111 (0.240)   &  \cite{Hua2020}   & 217 105 180.0 (1.1)  & 2.1 (1.1)  &  1.9 & T\\
  &  & 217 105 181.901 (0.076)  &  \cite{Castrillo2021b}   & 217 105 180.0 (1.1)  & 1.9 (1.1)  &  1.7& T \\
   & R(2) & 219 042 856.621 (0.025)  &  \cite{Cozijn2018}\footnote{Results from fitting center frequency of a Lamb-dip without considering the complex line shape.}   & 219 042 854.7 (1.1)  & 1.9 (1.1)  &  1.7 & T\\
& R(3) & 220 704 304.951 (0.028)   &  \cite{Cozijn2018}$^{\rm c}$  & 220 704 303.0 (1.1)  & 1.9 (1.1)  &  1.7 & T \\
&  & 220 704 304.984 (0.065)  & Present$^{\rm b}$ & 220 704 303.0 (1.1)  & 1.9 (1.1) & 1.7   & T \\
\hline
(2-2) \\
   & S(2)$^{\rm a}$ & 16 882 368.179 (0.020)  & Present  & 16 882 367.976 (0.140)  & 0.20 (0.14)   & 1.5  & T  \\
\hline
\end{tabular}
\end{table*}

In the present study the vibrational transitions R(1), R(3) and P(3) in the (2-0) overtone band of HD were measured in saturation via the NICE-OHMS technique.
These results are compiled in Table~\ref{tab:comparison} including all precision measurements on ro-vibrational and purely rotational transitions hitherto performed.
Older data on Doppler-broadened spectroscopy of vibrational overtone  transitions~\cite{Durie1960,Kassi2011,Vasilchenko2016} are not included.
While for the transition frequency of the R(1) line the result based on the systematic study of the line shape, at an accuracy of 50 kHz, was taken~\cite{Diouf2019}, results in the present study of R(3) and P(3)
are accurate to 65 and 60 kHz, respectively.
The experimental results are compared with values obtained via advanced ab initio calculations as in the H2SPECTRE program suite~\cite{SPECTRE2020}.
In this program some level energies and transitions are computed via non-adiabatic perturbation theory (NAPT)~\cite{Czachorowski2018}, while for some specific levels the non-relativistic part is computed via pre-Born-Oppenheimer or 4-particle variational calculations~\cite{Puchalski2018,Puchalski2019}.
The theory entries included in Table~\ref{tab:comparison} are partly based on the more accurately computed values, although some matrix elements and the Bethe-logarithm were computed on a BO-basis~\cite{Pachucki-private}.

Inspection of the Table shows that for the vibrational transitions there are a large number of entries marked by 'T', where the experimental values are more accurate than the theoretical ones and where the uncertainty is fully determined by theory.
Although the deviations are all in the range 1.6-1.9$\sigma$, or at 10 ppb, it is remarkable that the offsets are so consistently equal and of the same sign.
This may be considered as an indication that the ab initio calculations of H2SPECTRE  systematically underestimate the vibrational level spacings in HD.
Also in the  D$_2$ molecule a recent study yielded a similar underestimate of the theoretical value for the S(0) (1-0) vibrational ground tone frequency by 1.2$\sigma$~\cite{Fast2021}.
Also in that case, with an experimental accuracy of 17 kHz, the uncertainty was fully determined by theory~\cite{Komasa2019}.

For the measurements of pure rotational transitions performed so far there is only a single experimental result claiming the same accuracy as that of theory: a measurement of the R(0) line reported in Ref.~\cite{Drouin2011}. For this result the experimental value is again higher by 0.7$\sigma$ of the combined uncertainties.
The present measurements of rotational level spacings, deduced from combination differences of measured transition frequencies, represent equally accurate determinations, with their 20-30 kHz systematic uncertainties.

So when comparing the most accurate experimental data with the most advanced ab initio calculations, including relativistic and QED effects~\cite{Puchalski2016,Czachorowski2018,Komasa2019} the body of experimental data, both vibrational and rotational, are some 1.5-1.9$\sigma$ larger than theory.
This might be viewed as an offset scaling with energy.
However, that finding cannot be extrapolated to results on dissociation energies of the H$_2$ molecule, where the most accurate experimental result~\cite{Holsch2019} is found to be in excellent agreement with theory~\cite{Puchalski2018,Puchalski2019}.
A recent experimental value for the dissociation energy of the D$_2$ species~\cite{Hussels2022} was found to be off from theory~\cite{Puchalski2019} by 2 MHz, corresponding to 1.6$\sigma$,
but here the uncertainty contributions from experiment and theory were the same.
As for the case of the dissociation energy of the HD isotopologue the current experimental value~\cite{Sprecher2010} is off from theory~\cite{Puchalski2019} by 2.7$\sigma$.
While this might be viewed as another discrepancy for the HD species, where $g-u$ symmetry-breaking plays a role,  it should be considered however, that in this case the uncertainty is fully determined by experiment.

Inspection of results from the H2SPECTRE on-line program~\cite{SPECTRE2020} reveals that the uncertainties for the rotational splittings on the theoretical side fully depend on the uncertainty in the evaluation of the $E^{(5)}$ leading order QED-term.
For the $\Delta_{(J=3)-(J=1)}^{v=0}$ splitting the uncertainty in $E^{(5)}$
amounts to 107 kHz, compared to a full uncertainty over all terms of 110 kHz. For $\Delta_{(J=4)-(J=2)}^{v=2}$ this is 137 kHz out of 140 kHz uncertainty contributed from  $E^{(5)}$.
These uncertainties in the rotational splittings imply already strong cancellation of common-mode contributions, where the uncertainties in the $E^{(5)}$-term in the binding energies of HD-levels ($v=0,J=1$ and $3$) amount to 5.5 MHz~\cite{SPECTRE2020}.
For specific low-lying levels the $E^{(5)}$ contributions to their binding energy are much more accurate, like for the H$_2$ ($v=0,J=0$) ground level where $E^{(5)}$ is accurate to 5 kHz~\cite{Puchalski2019b},
while for HD ($v=0,J=0$) the  $E^{(5)}$-term is accurate only to 120 kHz~\cite{Puchalski2019}. This reflects the higher level of computation pursued for H$_2$, an approach that might resolve the presently found discrepancies between experiment and theory for the HD molecule.
In this sense the presently determined splittings in HD form a test bench theory for the further development of theory for the hydrogen molecular species.

\begin{acknowledgments}
The authors thank IHM van Stokkum (VUA) for fruitful discussions on the data analysis.
The research was funded via the Access Program of Laserlab-Europe (Grant Numbers 654148 and 871124), a European Union’s Horizon 2020 research and innovation programme.
Financial support from the  European Research Council (ERC-Advanced Grant No. 670168) and from the Netherlands Organisation for Scientific Research, via the Program “The Mysterious Size of the Proton” is gratefully acknowledged.
M. Schlösser wants to thank the Baden-Württemberg Foundation for the generous support of this work within the Elite-Postdoc-Fellowship.
\end{acknowledgments}

%

\end{document}